\documentclass[12pt,a4paper,twoside,final,notitlepage]{article}
\usepackage[dvips]{graphicx}
\usepackage[leqno]{amsmath}
\usepackage{a4}
\usepackage{epsf}

\usepackage{epsfig}
\usepackage{amsfonts}
\usepackage{amssymb}
\usepackage[english]{babel}
\usepackage[T1]{fontenc}
\usepackage{amsthm,bbm, amssymb,array,mathrsfs}
\usepackage{xspace}    
\newcommand{\pequationdeb}{$$ \left\{ \begin{minipage}[c]{130mm}}
\newcommand{\pequationfin}{\end{minipage}
                           \right. $$}

\newcommand{\beq}     {\begin{equation}}
\newcommand{\enq}     {\end{equation}}
\newcommand{\be}    {\begin{enumerate}}
\newcommand{\ee}    {\end{enumerate}}

\newcommand{\Bb}

\newcommand{\bbbc}{\mathbb{C}}
\newcommand{\bbbn}{\mathbb{N}}

\newcommand{\Sum}     {\displaystyle \sum} 
\def\resume{\if@twocolumn
\section*{R\'esum\'e}
\else \small
\quotation{\bf \it R\'esum\'e \rule[1mm]{1.5mm}{0.2mm}\vspace{0pt}}
\fi}
\def\endresume{\if@twocolumn\else\endquotation\fi}
\def\abstract{\if@twocolumn
\noindent\section*{{\bf Abstract}}
\else \small
\quotation{\noindent \bf {Abstract.} \rule[1mm]{1.5mm}{0.2mm}\vspace{0pt}}
\fi}
\def\endabstract{\if@twocolumn\else\endquotation\fi}

\begin{document}
\title{\bf \Large On Voting Process and Quantum Mechanics   ~\\~  }
\author { { \large  Fran\c{c}ois Dubois~$^{ab}$  }     \\  
{\it \small  $^a$  Conservatoire National des Arts et M\'etiers, Department of
  Mathematics,  Paris, France.}  \\
{\it  \small $^b$   Association Fran\c caise de Science des Syst\`emes. }   \\
{ \rm  \small duboisf@cnam.fr } ~\\  ~\\   }  
\date{ \normalsize 06 february 2009~\protect\footnote{~Presented  to 
{\it Quantum Interaction QI2009},  Saarbr\"ucken, Germany, 25-27 March 2009. 
Published in {\it Lecture Notes in Computer Science},  number 5494 (P. Bruza {\it et al}
Editors), p.~200-210, Springer, 2009. }}

\maketitle

~\\ \\

\noindent {\bf Abstract} \\ 
In this communication, we 
 propose a  tentative to set the fundamental problem 
of measuring  process done by a large structure on a microscopic one. 
We consider  the example
of  voting  when an entire society tries to measure globally opinions of all
social actors in order to elect a delegate. We present a quantum model to interpret an
operational voting system and propose an quantum approach for grading step of  
 Range Voting,  developed by  
M.~Balinski and R.~Laraki in 2007.

$ $ 

  \newpage
\section{ Measure process between different scales}
\noindent $\bullet$ \quad  
Matter is constituted by discrete quanta 
 and this fact was empirically put in
evidence by E.~Rutherford in the beginning of 20th century. 
Microscopic quanta as classical atoms or photons are not directly perceptible by our
senses, as pointed out by  M. Mugur-Sch\"achter \cite{MMS08}.
In consequence, any  possible knowledge for a human observer 
of a microscopic quantum  is founded    on  experimental protocols. 
The mathematical framework  constructed during the 
20th century  describes unitary ``free evolution'' through the Schr\"odinger equation 
and ``reduction of the wave packet'' associated to   measure process 
through a projection operator in  Hilbert space. We refer 
  the reader {\it e.g.} to the book
of  C. Cohen-Tannoudji {\it et al} \cite{CDL77}. 
The philosophical consequences of this new vision of Nature are still under  construction; 
in some sense, an {\it a priori} 
or an external  description of Nature is not possible at quantum scale. We 
  refer to   B. D'Espagnat \cite{Es02} and  M.~Bitbol \cite{Bi96}. 
Independently of the  development of this renewed physics, the importance of scale
invariance have been recognized by various authors as B.~Mandelbrot  \cite{Ma82} 
and L.~Nottale  \cite{No98}. The word ``fractal'' is devoted to figures and properties
that are self-similar whatever the refering  scale. 

  \smallskip \noindent $\bullet$ \quad
We have suggested in 2002  the fractaquantum hypothesis \cite{Du02}, founded on two remarks:
Nature develops a scale invariance  and quantum mechanics is completely relevant for
small scales. In order to express this hypothesis, we have introduced 
  (see {\it e.g.}  \cite{Du05, Du08a})  the notion of ``atom'',
   in fact very similar to the way of vision of Democrite and the
ancient Greek philosophers (see {\it e.g.} J.~Salem  \cite{Sa97}). 
To fix the ideas, an ``atom'' can be a classical atom, 
or its nucleus, or a molecule, or a micro-organism like a cell, or an entire
macro-organism as a human being or till an entire society! 
If we divide an ``atom'' into two parts, its qualitative properties change strongly at
least in one of these parts.
With this framework, elementary components are supposed to 
exist in Nature at different scales. A classical atom is a ``micro state'' relative to a
Human observer. 
In this particular case, 
a  $\ell$ittle  ``atom''  $\ell$ is a
 classical  atom and  a Big ``atom'' B is a human observer. 
More generally,  two  ``atoms'' $\ell$ and B have different scales when 
 ``atom'' $\ell$ is not directly perceptible to  ``atom'' B.   
In other words, a direct interaction between B and $\ell$ can not be controlled by B
himself.  In this case, the direct
interaction between little  ``atom'' $\ell$ and big  ``atom'' B can be neglected as a
first order approximation.

  \smallskip \noindent $\bullet$ \quad
In this contribution, we suggest to revisit this classical
 quantum formalism when little and big ``atoms'' are nonclassical ones. 
In fact, this research program is tremendous! 
For similar  programs, we   refer {\it e.g.}   to the works of  
G.~Vitiello  \cite{Vi01}, P.~Bruza {\it et al} \cite{BKNME08},
A.~Khrennikov and  E.~Haven \cite{KH07},   P.~La Mura  {\it et al} \cite{LMS07}.  
 The phenomenology of possible measurement
interactions should be reconstructed.
 What is a  big   ``atom'' B that can measure some
quantities on little   ``atom''  $\ell$? Does the classical framework of quantum mechanics
operates without any modification? Of course all these questions 
motivate our   communication.
Due to the lack of knowledge of what can be a measure done by ``atoms'' at mesoscopic 
or microscopic scales, we
restrict ourselves 
in this contribution 
to measures done by human society considered as a whole on individual  human beings.    
  
  \smallskip \noindent $\bullet$ \quad
We    consider  here   a   particular example of   the measurement
process associated with   voting.  
In this case,  ``atom'' $\ell$ is a social actor and  ``atom'' B is the entire society. 
We first introduce the scientific problem of   voting process and in 
 the following   section, we  present a preliminary quantum model for voting. 
In the two following sections we   describe with the help of 
fractaquantum hypothesis the range voting procedure (``vote par valeurs'') developed
independently by 
  M.~Balinski and R.~Laraki \cite{BL07a} at Ecole Polytechnique (Paris) 
and  by W.D.~Smith   \cite {Sm07, RS07}  at  the ``Center of   Range Voting'' (Stony Brook, New York). 

\section{On the voting process }    
\noindent $\bullet$ \quad  
 We consider a macroscopic ``atom'' B composed by an entire  social structure. 
For example, B is a state like France to fix the ideas. 
The social actors of  society B are the little
``atoms'' $\ell$ in our  model.  We  write here 
\begin{equation}
\label{ldB}
\ell \in {\rm B}  
\end{equation}
even if the   expression  (\ref{ldB})  
does not take precisely into account the detailed structure of society $ {\rm B}$. 
The numbers of such indistinguable individuals are  quite important ($10^6$ to $10^9$
typically). The democratic life in society B suppose that social responsabilities are
taken by elected representants of social corpus. 
Thus a voting process has the objective to determine
one particular social actor among all for accepting social responsabilities. This kind of
position is supposed to be attractive and a set $\Gamma$ of candidates $\gamma$  
among the entire set of    ``atoms'' $\ell$ is
supposed to be given in our framework. 
 
  \smallskip \noindent $\bullet$ \quad
The problem is to determine a single ``elected'' candidate $\gamma_1$ among the family
$\Gamma$ thanks to the synthesis of all opinions of different electors  $\ell$. 
  The social objective of society B is the determination
 of one candidate among
 others through a social process managed by the entire society, modelized   here  as
 a macro  ``atom'' B. 
This problem is highly ill posed and we refer to the pioneering  works of 
J.C. de~Borda  \cite{Bo1781} and N.~de~Condorcet \cite{Co1785} 
followed more recently by the theorem of 
non existence of a social welfare function satisfying reasonable hypotheses,  
proved by K.~Arrow  \cite{Ar51}. 
We describe this result in the following of this section.

  \smallskip \noindent $\bullet$ \quad
With K.~Arrow, we suppose that 
 each elector $ \, \ell \, $  determines  some ordering 
denoted by  $ \, \succ_{\sigma_{\ell}} \,$ (or simply by  $ \, \sigma_{\ell} $)
among the candidates $ \, \gamma \in \Gamma \, $ : 
\[
\gamma_{\sigma_l(1)} \,  \succ_{\sigma_{\ell}} \, 
\gamma_{\sigma_l(2)} \,  \succ_{\sigma_{\ell}} \, \dots \,   \, 
\gamma_{\sigma_l(i)} \,  \succ_{\sigma_{\ell}} \, \gamma_{\sigma_l(i+1)} \, 
 \,    \, \dots \,    \succ_{\sigma_{\ell}} \,\gamma_{\sigma_l(K)} \,, 
\quad  \ell \in {\rm B}.   
\]   
We consider now the set $\, \sigma \,$ of {\bf all} orderings $ \, \sigma_l \,$ for all
the electors $ \, \ell \,$ 
\[
\sigma \, = \, \{  \sigma_{\ell} ,\, \sigma_{\ell} \, \,  \textrm  {ordering of candidates}
\,  \Gamma,\, \ell \in B \} \, . 
\]   
A so-called social welfare function $ \, f \,$ determines a particular social ordering 
$ \,   \sigma^*= f(\sigma) \, $  
as a global synthesis of all orderings $\,  \sigma_{\ell} \,$ 
in order to construct  a commun and socially coherent position. 
Some democratic properties are {\it a priori} required for this function  $ \, f \, $: 

 (i)  \emph{  \textbf{ Unanimity}}

\noindent
If everybody thinks that candidate $\, \gamma \,$ is better than $ \, \gamma' \,$ the
social choice must satisfy this property:
\begin{equation}
\label{unanim} 
 {\rm If} \,\,\,   
\left( \,  \forall \ell  \in B ,\, \,\, \gamma    \succ_{\sigma_{\ell}}  \gamma' \,\right) \quad 
 \textrm  { for some } \, \gamma ,\,   \gamma'  \, \in \Gamma, \quad 
{\rm then} \, \, \left(  \gamma   \succ_{\sigma^*}  \! \gamma' \, \right)  .  
\end{equation}

 (ii)  \emph{  \textbf{ Independance of irrelevant alternatives }} 

\noindent Consider two orderings $ \,\sigma  \, $ and   $ \,\tau  \, $ 
 grading in a similar way the two candidates  $\, \gamma \,$ and $ \, \gamma' \, $: 
\begin{equation}
\label{indep-1} 
   \left( \left( 
  \gamma    \succ_{\sigma_{\ell}}  \! \gamma' \right)  \,
 {\rm and}  \left( \gamma    \succ_{\tau_{\ell}}  \! \gamma' \right) 
\right) \,\,  {\rm or }  \,\, 
 \left(  \left( \gamma    \prec_{\sigma_{\ell}}  \!  \gamma' \right)  \,
 {\rm and}  \left( \gamma    \prec_{\tau_{\ell}}  \! \gamma' \right) 
\right)    \,, \quad   \forall \ell  \in B \, .  
\end{equation}
Then the social orderings $ \, \sigma^* = f(\sigma) \,$ and 
 $ \, \tau^* = f(\sigma) \,$ must satisfy the corresponding property:
\begin{equation}
\label{indep-2} 
 \gamma  \succ_{\sigma^*} \!  \gamma \,\,\, {\rm when}  \left( \left( 
  \gamma    \succ_{\sigma_{\ell}}  \! \gamma' \right)  \,
 {\rm and}  \left( \gamma    \succ_{\tau_{\ell}}  \! \gamma' \right)  \right)  
\,\,   {\rm or }  \,\, 
 \gamma  \! \prec_{\sigma^*} \!  \gamma \,\,\, {\rm when}  \left( \left( 
  \gamma    \! \prec_{\sigma_{\ell}}  \! \gamma' \right)  \,
 {\rm and}  \left( \gamma    \! \prec_{\tau_{\ell}}  \! \gamma' \right)  \right) \, .   
\end{equation}
The social welfare function depends only on the  relative ranking and not on the
intermediate candidates. 
 
  \smallskip \noindent $\bullet$ \quad
Then the Arrow impossibility theorem (proven elegantly 
 by  J.~Geanakoplos in \cite{Ge01}) 
implies that under conditions   (\ref{unanim})  of unanimity 
and (\ref{indep-1})-(\ref{indep-2})  of independance of irrelevant alternatives, 
the social welfare function is simply a constant:

 (iii)  \emph{  \textbf{ Dictatorship }} 
\begin{equation}
\label{dictat} 
\exists \, d \in \Gamma \,, \quad  f( \{ \sigma_{\ell},\, \ell \in B \} ) \,\equiv \, \sigma_d \, 
\end{equation}
and the result is a dictature! 
In other terms, it is impossible to construct a social welfare function that has the two
first properties of  unanimity  and  independance of irrelevant alternatives
and the non-dictatorship property, obtained by negation of (\ref{dictat}).

\section{A preliminary quantum model for voting} 
 \noindent $\bullet$ \quad
We describe in this Section a quantum model presented in \cite{Du08b}. 
We restrict here to the so-called ``first tour'' process as implemented 
in a lot of situations. 
In this process, each elector  $\ell$ has to transmit  the
name of   at most  {\bf one} candidate $\gamma$. 
Then an ordered list of candidates is obtained by counting
the number of expressed votes for each candidate. 
 Introduce the space $H_{\Gamma}$ of candidates 
generated formally by the finite family $\Gamma$ of all candidates: 
\begin{equation}
\label{HG-space}
H_{\Gamma} \,= \,\bigoplus_{\gamma \in \Gamma} \,\, \bbbc   \,   | \, \gamma \! >      
\end{equation}
where $\,  \bbbc\, $ denotes the field of complex numbers. 
This decomposition (\ref{HG-space})   is supposed to be orthogonal: 
\[  
 <  \gamma \, | \, \gamma' \! >   \,= \, 
\left\{
\begin{array}{rcl}
0 & {\rm if}  & \gamma \not =  \gamma'    \\  [1mm] 
1 & {\rm if}  & \gamma  =  \gamma',      
\end{array}
\right. \, , \qquad \gamma, \gamma'  \in \Gamma. 
\] 
The ``wave function'' associated with an elector  $\ell$ is represented by a state 
 denoted by   $\, |  \, \ell \! > \, $ in   this space  $H_{\Gamma}$:
\begin{equation}
\label{elec}
 |  \, \ell \! > \,= \, \sum_{ \gamma  \in \Gamma} \, | \,\gamma \! > 
 \, < \ell \, | \,\gamma \! >  \,  \, . 
\end{equation}  
The scalar product $\, <  \ell \, | \, \gamma \! >  \,$  in relation  (\ref{elec}) 
is the component of elector  $\ell$ relative to each candidate $\gamma$. This number
             represents the  political sympathy
of elector  $\ell$ relative  to the candidate $\gamma$. We suppose here that the norm 
$\, \parallel \! \ell \! \parallel \,$ of state   $\, |  \, \ell \! > \,  $ {\it id est}
\[  
\label{norm}
 \parallel \! \ell \! \parallel  \,\equiv  \, \sqrt{  \sum_{ \gamma  \in \Gamma} 
\mid   <  \ell \, | \, \gamma \! >     \mid ^2 \, } 
\]   
is {\bf inferior or equal} to unity. We follow the Born rule and suggest that the
probability for elector $\ell$ to give its vote to candidate $\gamma$ is equal to 
$\, \mid    <  \ell \, | \, \gamma \! >    \mid ^2 .\,$ 
We suggest also  that the   probability to unswer by a vote ``blank or null'' is 
$ \,\, 1 -  \parallel \! \ell \! \parallel^2 \,\, $ in this framework.

  \smallskip \noindent $\bullet$ \quad
The interpretation of the projection process in the quantum measurement for such a first
tour of election process is quite clear. During  the election, {\it id est}   the particular 
day where  the measure process occurs, the elector   $\ell$  is  {\bf obliged }
to choose  at most one candidate $\gamma_0$. In consequence, 
 all his political sensibility is socially ``reduced'' to this particular  candidate.
We can write: 
\[ 
\label{after}    \, |  \, \ell \! >   \,= \,   \, |   \gamma_0  \! >  \, 
\]    
to express the wave function collapse.
This quantum interpretation of such  voting  process clearly shows the  {\bf violence}
of such king of decision making. Of course, no elector has political opinions that are
identical to one precise candidate and this measurement process is a true
mathematical  projection. 
Nevertheless, the operational social voting  process imposes this projection in order to
construct a social choice. The disadvantage and dangers of such process have been
clearly  demonstrated in France during the presidential election process in 2002  
(see {\it e.g.}  \cite{Fr02}).
 
\section{Range Voting (i): quantum approach for grading step} 
\noindent $\bullet$ \quad   
The voting process suggested by  M.~Balinski and  R.~Laraki  \cite{BL07a} 
is more complex than the one studied in the previous section. 
The key point in order to overcome the Arrow impossibility theorem 
is the fact that in this framework the opinion of electors among the candidates are
{\bf  codified}  by society B
through a given set of so-called ``grades''.  
These grades are {\it a priori} very similar to the ones given by the scolar  system, as
integers between 0 and 20 in France with an associated order 
\[     
0 \,\prec\,  1  \,\prec\,  \dots  \,\prec\,  j  \,\prec\,  j+1  \,\prec\,   \dots
  \,\prec\,    19  \,\prec\,  20 \,, 
\]     
letters   from A to F in the United States 
with an order   
\[      
{\rm A}  \, \succ \,   {\rm B}  \, \succ \,   {\rm C}  \, \succ \,   {\rm D}  
\, \succ  \,   {\rm E}  \, \succ \,   {\rm F} \,, 
\]     
or numbers from 1 to 6 in Germany with the following (mathematically unusual!) order 
\[      
 1   \, \succ \, 2   \, \succ \,  3    \, \succ \,   4   \, \succ \, 
 5    \, \succ \,  6 \, . 
\]     
These grades can be also  an ordered  list of given words 

\smallskip \noindent 
``very  good''  $\, \succ \,$ 
``good''        $\, \succ \,$  
``not   so bad''      $\, \succ \,$  
%
   ``passable''   $\, \succ \,$  
``insufficient'' $\, \succ \,$  
``to  be  rejected'' 

\smallskip \noindent   
as proposed by the previous authors \cite{BL07b} 
in  Orsay  experiment for French  presidential election in 2007.
These grades define an elementary {\bf  common language } that is supposed to be 
endowed by all social actors $\ell$ of society B. In other terms,   
 a common ordered set $G$ of grades $\nu$ is supposed to be given: 
\begin{equation}
\label{notes} 
\nu_1  \,\succ \, \nu_2   \,\succ  \,  \dots  \,\succ \,  \nu_K \,, \quad \nu_j \in G \, .   
\end{equation}
As a consequence, an ordering of opinions explicitly refer to 
this particular set of given grades and to an
explicit  ordering between these grades like in (\ref{notes}). 
Remind  that in Balinski-Laraki 
process \cite{BL07a}, the society B imposes  a commun grading  referential to all
electors.

  \smallskip \noindent $\bullet$ \quad
The ranking process between the candidates proceeds by two steps. First each elector
gives a grade to each candidate.  
Secondly the candidates are arranged in order through ``majority ranking''. 
Each elector $\ell$ has to express an opinion relative to each candidate 
$\, \gamma \in \Gamma \,$ through a grade $\, g(\gamma,\, \ell) \in G. \,$ 
During the day of the election as in \cite{BL07b}, each elector grades each candidate. 
We  propose in this section a quantum model for the first step of this processus.  
This first step
is a measure done by society B on each little ``atom''  $\ell$ which constitutes
it, as suggested by relation (\ref {ldB}). 
Observe now that each   candidate $\gamma$ has a published political
program, is giving radio and  television interviews, has a blog, {\it etc.} 
  We introduce a ``political  Hilbert space''  
$\, H_P \,$ that  refer to {\bf all} this set of political information, 
following modern approaches for 
   Information Retrieval as suggested by K.~von~Rijsbergen \cite {vR04}. 
The family $G$ of grades is imposed by the general laws of society B.
Nevertheless, the evaluation  of the political program of all 
candidates  is done by the elector $\ell$ himself in such a process! 
We suggest that each elector $\ell$ decomposes this Hilbert space  $\, H_P \,$ 
 into ``grading'' orthogonal components $\, E^{\ell}_{\nu} \,$ through 
his  own   {\bf internal}  
  process: 
\begin{equation}
\label{decomposition} 
H_P \,= \,\bigoplus_{\nu \in G} \,  E^{\ell}_{\nu} \, , \quad \ell \in {\rm B} \, .  
\end{equation}
The subspace  $\, E^{\ell}_{\nu} \,$ is the eigenspace giving the grade $ \nu \,$ 
relative to the opinion of elector~$\ell$. If we
 denote by $A^{\ell}$ the quantum self-adjoint operator associated with 
the grading process done by  elector $\ell$, we  have 
\begin{equation}
\label{vect-propre}  
A^{\ell}  \,  {{\scriptstyle \bullet }} \, |  \, \xi \! >  \, = \, \nu \,  
 \, |  \, \xi \! >  \,,\quad    \, |  \, \xi \! > \, \in  E^{\ell}_{\nu} 
\,\subset \,  H_P  \,  \, , \quad \nu \in G \, . 
\end{equation} 
In other words,  we introduce the  orthogonal projector $ \ P^{\ell}_{\nu} \,$  onto the
closed   space $ \, E^{\ell}_{\nu} $. Then these projectors commute 
\[
 P^{\ell}_{\nu} \,\,  P^{\ell}_{\nu'} \,\, = P^{\ell}_{\nu'} \,\,  P^{\ell}_{\nu} \, \,,
 \qquad \nu \,, \,   \nu' \, \in G \,, \,\,\, \ell \in {\rm B} \,  
\]
and generate 
a decomposition of the identity operator  $ \, {\rm Id}(H_P) \,$  in the political 
Hilbert space $ \, H_P \,$: 
\begin{equation}
\label{projecteurs} 
\Sum_{\nu \in G} \,  P^{\ell}_{\nu} \, \equiv \,  {\rm Id}(H_P)  \, , \quad \ell \in {\rm B} \, .  
\end{equation}
On a very concrete point of view, in front of each political idea, each  elector has the
capability to give an opinion  in the language suggested  {\it a priori} by 
 the set $ \, G \, $ of grades.  The examples of such sets 
given above show also that the way of  decomposition of political space $ \, H_P \,$
through the grades is strongly influenced by the social choice of the family $\, G .$

  \smallskip \noindent $\bullet$ \quad
In some sense,  {\it via} a particular choice of grading, the society B
imposes some filtering of space $ \, H_P \,$ of all political data. 
Note that the precise way this filter is done depends on each citizen $ \, \ell . $ 
In this model,  society B imposes the set $G$ of eigenvalues   
and each elector $\ell$ fixes the eigenvectors as in (\ref{vect-propre}).  
After the elector has interpreted the grades $\, \nu \,$ in his own vocabulary, {\it id
  est} once he has decomposed the space  $H_P$ into orthogonal components, 
we suppose that the grading process, {\it id est} 
the result of the measure is {\it a priori} obtained according to the
  Born rule.  
Precisely, we introduce  the  ``perception''   $ \, \rho^{\ell}_{\gamma} \,$ 
of political opinion of candidate $ \, \gamma \,$ by the elector
$  \, \ell .$ Mathematically speaking, the elector $  \, \ell \, $ measurates
the political ideas of the  candidate 
  $ \, \gamma \,$ in a quantum way relatively to   the Hilbert space $ \, H_P .$ 
According to  Gleason theorem \cite{Gl57}, such a quantum probability 
is defined by a density matrix, {\it id est} 
a positive self-adjoint operator of unity-trace  that we denotes also by  
$ \, \rho^{\ell}_{\gamma} \,$: 
\[
 \rho^{\ell}_{\gamma} \,\, \textrm  {positive self-adjoint operator} \,\, H_P \,
 \longrightarrow  \,  H_P \,, \quad {\rm tr} \, \left( \rho^{\ell}_{\gamma}   \right) =  1 \, . 
\]
Then, following A.~Gleason  \cite{Gl57} and  K.~von~Rijsbergen \cite {vR04}, 
 the measure $ \, \mu^{\ell}_{\gamma} \,$  
associated with  elector $ \, \ell \,$ and candidate $ \, \gamma \,$  
of  any closed subspace  $ \, E \subset H_P \,$ 
is given in all generality according to 
 \begin{equation}
\label{proba} 
\mu^{\ell}_{\gamma} (E) \,=\, {\rm tr} \, \left( \rho^{\ell}_{\gamma} \, P_E \right)   \,, 
\quad E \subset H_P\,, \,\,\, \ell \in {\rm B} \,, 
\end{equation}
where $\, P_E \,$ is the orthogonal projector onto  space $\, E .$ 
Consider now the space $ \, E = E^{\ell}_{\nu}  \, $ introduced in (\ref{decomposition}).
Then the (real!) number $ \, \mu^{\ell}_{\gamma, \nu}  \, $ defined by 
 \begin{equation}
\label{proba-2} 
 \mu^{\ell}_{\gamma, \nu} \,=\,  \mu^{\ell}_{\gamma} ( E^{\ell}_{\nu}) \,=\, 
{\rm tr} \, \left( \rho^{\ell}_{\gamma} \, P^{\ell}_{\nu}  \right)   \, 
\end{equation}
represents the quantum probability for elector $ \, \ell \,$ to give the grade $ \, \nu \,$ 
to   candidate $ \, \gamma . \,$ Of course, if we insert 
the identity operator $ \,  {\rm Id}(H_P) \,$ 
decomposed in (\ref{projecteurs}) inside relation (\ref{proba}), we have due to  (\ref{proba-2})
\begin{equation}
\label{coherent} 
\Sum_{\nu \in G} \,   \mu^{\ell}_{\gamma, \nu} \, =\, 1 \,, \qquad  \ell \in {\rm B},\,\, 
\gamma \in \Gamma ,   
 \end{equation}
and the sum of probabilities for all  different grades is equal to unity. 

 \smallskip \noindent $\bullet$ \quad
Remark  that two different ingredients are necessary to determine the previous probability 
$ \,  \mu^{\ell}_{\gamma, \nu} \,$ in (\ref{proba-2}). First the decomposition 
(\ref{decomposition}) of the political space through the grades $ \, G .$ 
As usual in quantum mechanics, no detailed structure of ``atom'' $\ell$ is transmitted
through the measure process. In this case, the orthogonal decomposition 
   (\ref{decomposition}) is not known by the society.   
%
Second the
``perception operator''  $ \, \rho^{\ell}_{\gamma} \,$ which represents in some sense the 
particular ``political knowledge''  that the elector $ \, \ell \, $ has constructed for
himself about the candidate $ \, \gamma . \,$  
Remark that no direct interaction between the candidates occurs in the model. 
According to Condorcet's ideas \cite{Co1795}, each  citizen is adult has make his own opinion through 
his own way of thinking! 
%

\section{Range Voting (ii): majority ranking } 
 \noindent $\bullet$ \quad 
After this first step of grading, the result of the vote of elector
$\ell$ is a list 
\[       
\label{vote}
g(\gamma,\, \ell) \, \in G \,, \quad \gamma \in \Gamma   \,, \quad \ell \in {\rm B} 
\]       
of grades $\nu  = g(\gamma,\, \ell) \,  $ 
given by elector $\ell$ to {\bf each } candidate $\gamma$. 
We give in this section the major points introduced By Balinski and Laraki \cite{BL07a}
without any modification.
After summation, each candidate $\gamma$ has a certain number 
$\, n^{\gamma}_{\nu} \in \bbbn \, $ 
of opinions transmitted by the electors:  
\begin{equation}
\label{list} 
n^{\gamma}_{\nu} \, = \, \textrm { Card } \left\{  \, \ell \in B \,, \,\, g(\gamma,\, \ell) =
\nu \, \right\} \,  \in \bbbn  \,, \quad \gamma \in \Gamma \,, \quad \nu \in G.  
\end{equation}
The way of ranking  such a list 
\begin{equation}
\label{list-2} 
n^{\gamma} \,\equiv\, \big( n^{\gamma}_{\nu_1} \,, \, n^{\gamma}_{\nu_2} \,, \, \dots 
n^{\gamma}_{\nu_K} \big)   \in  \bbbn^K \,, \quad \gamma \in \Gamma 
\end{equation}
when the grades $\, \nu \in G \, $ are arranged in order without ambiguity by (\ref{notes})  
can be explicited with the so-called ``majority ranking''  introduced  
by Balinski and Laraki  \cite{BL07a}. 
 We give here some details   of the algorithm, based on a successive   extraction
of a {\bf median value}  from a list as  the one described  in  (\ref{list-2}) and  refer to
\cite{BL07a}, \cite{BL07b} and \cite{PB06}. 

  \smallskip \noindent $\bullet$ \quad 
From an algorithmic point of view, the list 
$ \, n^{\gamma} \, $ can also be written  as a list $ \, m^{\gamma} \,$ of  
grades written in decreasing order to fix the ideas: 
\begin{equation}
\label{list-3} 
m^{\gamma} \, = \,   \Big( \, \underbrace { \nu_1 ,\, \nu_1  ,\, \dots ,\, \nu_1
}_{ \displaystyle n^{\gamma}_{\nu_1} \, {\rm times} } \,,\,\, 
\underbrace { \nu_2 ,\, \nu_2  ,\, \dots ,\, \nu_2
}_{ \displaystyle n^{\gamma}_{\nu_2} \, {\rm times} } \,, \, \dots \,, \,\, 
\underbrace { \nu_K ,\, \nu_K  ,\, \dots ,\, \nu_K
}_{ \displaystyle n^{\gamma}_{\nu_K} \, {\rm times} } \,  \Big) \, 
\in \bbbn^{ \mid  B     \mid}
\end{equation}
where $ \,  \mid  \! B \! \mid  = {\rm Card}(B) \,$ is the number of electors. Then a list 
$ \, m^{\gamma}_1 \,$  can be constructed   by omitting the grade $ \, \nu^{\gamma}_{j_1}  \, $ 
at the  {\bf median } position
  $ \,  \frac { \mid  B \mid}{2} \,$ inside the list (\ref{list-3}).  
We obtain in this way a new list extracted from (\ref{list-3})
\begin{equation}
\label{list-4} 
m^{\gamma}_1 \, = \,   \Big( \, \underbrace { \nu_1 ,\, \nu_1  ,\, \dots ,\, \nu_1
}_{ \displaystyle n^{\gamma}_{1,\, \nu_1}  \, {\rm times} } \,,\,\, 
\underbrace { \nu_2 ,\, \nu_2  ,\, \dots ,\, \nu_2
}_{ \displaystyle n^{\gamma}_{1,\,\nu_2} \, {\rm times} } \,, \, \dots \,, \,\, 
\underbrace { \nu_K ,\, \nu_K  ,\, \dots ,\, \nu_K
}_{ \displaystyle n^{\gamma}_{1,\, \nu_K} \, {\rm times} } \,  \Big) \, 
\in \bbbn^{ \mid  B     \mid - 1 }
\end{equation}
and the integers  $ \, n^{\gamma}_{1,\, \nu_i} \,$ are equal to the  $ \, n^{\gamma}_{\nu_i} \,$
except for index $ \, j_1 \,$ for which we have 
\[
 n^{\gamma}_{1,\, \nu_{j_1}^{\gamma}} \,= \,  n^{\gamma}_{\nu_{j_1}^{\gamma}} - 1 \, . 
\]
The grade $ \,  \nu^{\gamma}_{j_1}  \, $ is the first ``majority grade'' 
 of candidate $ \, \gamma \,$ in the majority ranking algorithm of Balinski and Laraki. 
If  $ \,  \nu^{\gamma}_{j_1}   \succ  \nu^{\gamma'}_{j_1}  \, $
then we have the relative final  position    $ \,  \gamma     \succ \!  \gamma'  \, $ 
between the candidates $ \, \gamma \,$ and   
 $ \, \gamma' \,$.  If  $ \,  \nu^{\gamma}_{j_1} =  \nu^{\gamma'}_{j_1}  \, $
we apply the same step from (\ref{list-3})  to (\ref{list-4}) except that we start with
the list     (\ref{list-4}). Doing this, we extract a second grade  
$ \, \nu_{j_2}^{\gamma} \,$ for each candidate $\, \gamma \, $. 
If  $ \,  \nu^{\gamma}_{j_2}   \succ  \nu^{\gamma'}_{j_2}  \, $ or 
   $ \,  \nu^{\gamma}_{j_2}   \prec  \nu^{\gamma'}_{j_2}  ,\, $
the conclusion is established. Otherwise the process is carried on until 
the two majority grades at a certain step  are distinct.

  \smallskip \noindent $\bullet$ \quad
   It is a main  contribution of M.~Balinski and R.~Laraki \cite{BL07a} 
 to extract an intrinsic order  
\[       
\label{classement} 
 \gamma_1 \,\succ \, \gamma_2  \,\succ \, \dots  \gamma_j  \,\succ \, \gamma_{j+1}
 \,\succ \, \dots  \,, \quad \gamma_j \in \Gamma 
\]
among the candidates $\, \Gamma \, $ from the given double list   (\ref{list-2}) of  integers
$\, n^{\gamma} . $  
The important social fact is that  the overdetermination of a favorite candidate 
essentially does {\bf not} influence 
the final majoritary ranking with this grading method! 
The proof of this important fact is omitted here and we refer to  \cite{BL07a}. 
 We could also think that there is a contradiction between this positive result 
and the Arrow impossibility  theorem. In fact, as pointed in  \cite{BL07a}, the hypotheses
of Arrow theorem are qualitative: each elector consider some ordering of the candidates
with his own sensibility. As we have intensively explained with the orthogonal
decomposition (\ref{decomposition}), the social choice of a {\bf given} family of grades
is essential for the grading step and the majority ranking.

\section{Conclusion} 
\noindent $\bullet$ \quad 
The very   elaborated process initialized  by M.~Balinski and R.~Laraki   \cite{BL07a}
for range voting 
has been studied in this contribution. The second step of  ``majority ranking''
has been described without adding any new idea to this beautiful article. 
Concerning the first step of the algorithm devoted to the grading of each candidate 
by each elector with a given list of grades,  
 we have proposed a quantum algorithm essentially based on 
modern quantum approaches for Information Retrieval  presented  in K.~von~Rijsbergen's 
book \cite{vR04}. 
First an orthogonal  decomposition of the political Hilbert space 
supposes that each elector has the capability to have a precise opinion 
for each political subject.  
Second, following Gleason theorem  \cite{Gl57},  we have introduced a  
``perception operator'' that describes  mathematically the way a given candidate
is politically understood by a given elector. In some sense, a 
 psychological model is incorporated with  this description. 

\smallskip \noindent $\bullet$ \quad 
With these two ingredients, the computation of the probability for an elector to give 
a particular  grade to each candidate can be evaluated as a result of the model. 
Of course,  it is not actually clear which precise practical 
advantages has this quantum approach in the description of the voting process. 
Moreover, we want to   find in future works some previsions of the quantum model, 
and try to compare it with the previsions of a  classic model.

\noindent $\bullet$ \quad  
In this contribution, we have also presented a first quantum model of a  
classical election. In this framework, the big scale (the society) imposes  
 a direct generalization of the measure process in quantum mechanics.
All the characteristics of the mathematical measure operator
  are controlled by the large scale.  We have noticed the  violence 
 of the  multiscale interaction through such a the measuring process.   

\smallskip \noindent $\bullet$ \quad 
Last but not least,   this work is 
motivated by  the fractaquatum hypothesis \cite{Du02}. 
The case of a voting process is an example of measuring process between two different
scales in Nature. 
If we suppose that  the general concepts  of quantum
mechanics  have an  extension to all ``atoms'' in Nature, 
 the process of measuring has to be re-visited to all pairs 
of ``atoms''  with  different scales. 
This contribution is a small step in this direction!

\section*{Acknowlegments} 
The author thanks the referees who pointed clearly the importance of Information Retrieval 
 framework for this work and proposed a list of very interesting remarks 
and an important number of which have been incorporated in  the present writing!

\end{document}